\documentclass{article}

\usepackage{arxiv}

\usepackage[utf8]{inputenc} % allow utf-8 input

\usepackage{amsfonts}       % blackboard math symbols
\usepackage{nicefrac}       % compact symbols for 1/2, etc.
\usepackage{microtype}      % microtypography
\usepackage{lipsum}

\usepackage{color}
\usepackage{listings}
\usepackage{booktabs}
\usepackage{array}
\usepackage{graphicx}
\usepackage{longtable}

\usepackage{cite}
\usepackage{url}
\usepackage{fancyhdr}

\usepackage{soul}

\usepackage{float}
\usepackage{listings}
\usepackage{mdwmath}
\usepackage{mdwtab}
\usepackage{multirow}
\usepackage{multicol}
\usepackage{rotating}
\usepackage{setspace}
\usepackage[utf8]{inputenc}
\usepackage{lineno}
\usepackage{listings}

\usepackage{mdwmath}
\usepackage{mdwtab}
\usepackage{multirow}
\usepackage{multicol}
\usepackage{array}
\usepackage{booktabs}

\usepackage{enumitem}
\usepackage{xspace}
\usepackage[export]{adjustbox}
\usepackage{graphicx}
\usepackage{color,soul}
\usepackage{rotating}
\usepackage{setspace}
\usepackage{amsmath} 
\usepackage{amssymb}
\usepackage{float}
\usepackage{xcolor}

\usepackage{hyperref}
\usepackage[numbers]{natbib}
\usepackage{url}

\title{Defining Blockchain Governance Principles: A Comprehensive Framework}
\author{Yue Liu, Qinghua Lu, Guangsheng Yu, Hye-Young Paik, Liming Zhu\\
Data61, CSIRO, Australia\\
University of New South Wales, Australia\\
yue.liu@data61.csiro.au, qinghua.lu@data61.csiro.au\\
saber.yu@data61.csiro.au, h.paik@unsw.edu.au, liming.zhu@data61.csiro.au}
\begin{document}

\maketitle

\begin{abstract}
Blockchain eliminates the need for trusted third-party intermediaries in business by enabling decentralised architecture design in software applications. However, the vulnerabilities in on-chain autonomous decision-makings and cumbersome off-chain coordination lead to serious concerns about blockchain's ability to behave in a trustworthy and efficient way. Blockchain governance has received considerable attention to support the decision-making process during the use and evolution of blockchain. Nevertheless, the conventional governance frameworks {do not apply to} blockchain due to its distributed architecture and decentralised decision process. {These inherent features} lead to the absence of {a} clear source of authority in blockchain ecosystem. Currently, there is a lack of systematic guidance on the governance of blockchain. Therefore, in this paper, we present a comprehensive blockchain governance framework, which elucidates an integrated view of the degree of decentralisation, decision rights, incentives, accountability, ecosystem, and legal and ethical responsibilities. The above aspects are formulated as six high-level principles for blockchain governance. We demonstrate {a} qualitative analysis of the proposed framework, including case studies on five extant blockchain platforms, and comparison with existing blockchain governance frameworks. The results show that our proposed framework is feasible and applicable in a real-world context. 

\end{abstract}

blockchain, governance, decision right, accountability, incentive, ecosystem

\section{Introduction}
\label{sec:intro}
Blockchain is the technology behind Bitcoin~\cite{Satoshi:bitcoin}, which is a decentralised data store that maintains all historical transactions of Bitcoin network. The concepts of blockchain have been generalised to distributed ledger systems that ensure distributed trust without the need for third-party intermediaries in business transactions~\cite{scheuermann2015iacr}. A large number of projects have been conducted to explore how to re-architect application system. In particular, after the advent of smart contracts (i.e., computer programs run on blockchain), blockchain is explored as a general, decentralised computing and storage environment in which to build new applications and business models~\cite{2019-Bratanova-ACS}.
%In addition to the huge impact blockchain has brought to financial sectors, it provides a decentralised computing and storage infrastructure to build trustworthy decentralised applications (DApp). A plenty of studies have been conducted to explore the integration of blockchain in existing software systems, to enable decentralisation of trust without central authorities~\cite{2019-Bratanova-ACS}.
%Blockchain was designed as the underlying technology behind Bitcoin~\cite{Satoshi:bitcoin} and other following cryptocurrencies. In addition to the huge impact blockchain has brought to financial sectors, it provides decentralised computing and storage infrastructure to build trustworthy decentralised applications (DApp). A plenty of studies have been conducted to explore the integration of blockchain in existing software systems, to enable decentralisation of trust without central authorities~\cite{ukreport,aureport}.

Despite the broad use of blockchain, there are serious concerns about whether the decisions directing a blockchain are made in a trustworthy and efficient manner. A few
%The several 
infamous incidents occurred in two of the world-renowned blockchain platforms, Ethereum and Bitcoin, raise the level of concerns. %about blockchain decision making processes. %Blockchain platforms enable autonomous on-chain decision-making process by implementing algorithmic mechanisms to realise the concept of ``code is law"~\cite{selected19}. However, 
In 2016, within Ethereum, vulnerable code in smart contracts responsible for operating a DAO (Decentralised Autonomous Organisation) project was exploited by malicious attackers, which led to massive economic loss. After much debate, Ethereum conducted a hard fork to reverse the transactions in the DAO attack, and recover the stolen tokens worth over 60 million dollars~\cite{DAOattack}. Bitcoin also suffered {a split of the platform} after a lengthy debate over the block size (from August 2015 to January 2016)~\cite{selected20}. 
These incidents highlight the need for governance mechanisms that would orchestrate a clear decision-making process within the context of a decentralised system with different stakeholders. 
For instance, blockchain platforms need to be updated to fix software bugs without causing a hard fork, or different stakeholders need to be coordinated effectively and efficiently to come to a consensus while still respecting the principle of decentralised decision-making processes.
Blockchain governance refers to structures and processes which ensure that the development and use of blockchain are compliant {with} legal regulations and ethical responsibilities~\cite{liu2021systematic}. This topic has received great attention as it is considered essential to improve the trustworthiness and efficiency of blockchain. Nonetheless, existing IT/data governance frameworks and standards can hardly be applied to blockchain, as there is no explicit declaration of a central source of authority within blockchain. Recently, there are some works that discuss governance structures for blockchain platforms, either focusing on customised governance methods in permissioned blockchain~\cite{selected1, selected14}, or investigating regulations about financial sectors~\cite{selected3, selected16}. Some studies also propose governance frameworks for blockchain~\cite{selected11, selected14, hofman2021blockchain}. However, existing blockchain governance frameworks mainly discuss scattered governance mechanisms lacking stakeholders and process related linkages, which provide limited guidance to relevant stakeholders and {the} broader community who are interested in this topic.
%In particular,  can be hardly adopted by % The abstract discussion of technical implementation result in inadequate references for
%practitioners during the development and use of blockchain.}% to assess existing blockchain governance structures and guidance on future design and realisation.

%Nevertheless, previous studies pay little attention to the blockchain development process, which may result in the inadequacy of significant steps and involving stakeholders. Further, the abstract discussion of technical implementation exposes the lack of reference for practitioners to assess existing blockchain governance structures and guidance on future design and realisation.

%a specific governance method, or studying governance framework to guide the design directions of blockchain. Nevertheless, there are two critical weaknesses of existing works. On one hand, the lack of consideration on the different business context and concepts in the blockchain ecosystem result in the neglect of decentrlaisation level and governance structure of a blockchain platform. On the other hand, blockchain governance is an on-going topic comprising of multiple procedures, the overlook of a complete blockchain development process may lead to the inadequacy of significant steps and involving participants.

% There are...[related work]
% However, the existing frameworks largely contains disjointed governance mechanisms, lacking stakeholders and process related linkages.

Therefore, in this paper, we present a comprehensive blockchain governance framework that provides an integrated view of decentralisation level, decision rights, incentives, accountability, ecosystem, legal compliance and ethical responsibilities. % consists of six high-level governance principles taking into account the unique features of blockchain.%with comprehension over the features of blockchain.
%The framework can help researchers and practitioners to understand how governance is managed and implemented, making both governance decisions and related process more reliable and trustworthy. 
The major contributions of this paper are as follows:
%\helen{Would you have a look at this points again? I think the first two points can be combined into one - to highlight that you have look at existing governance structures extensively - and come up with the categories and mappings. The next two points can be combined into one - to say your framework extends the conventional scope of governance. Then have a new point to say what your framework is used for (i.e., guidelines???)}
\begin{itemize}

% \item  The proposed a blockchain governance framework categorises the adopted governance structure according to the level of decentralisation for the governed blockchain.

% \item The framework provides a systematic mapping of blockchain stakeholders and their respective decision rights, accountability, and incentives. This mapping can help inform stakeholders and the broader public about the authority, capability, and responsibility in blockchain governance.

% \item The framework extends governance to the blockchain ecosystem, which is comprised of four layers: data, platform, application, and community. We identify a set of governance mechanisms and associate each of them with the respective stage in the lifecycle at each layer.

% % linked with the associated governance mechanisms.%refined governance methods and mechanisms. To our best knowledge, this is the first blockchain governance framework with fine-grained development process.

% \item The framework emphasises legal compliance and ethical responsibilities in blockchain governance. We analyse the integration of existing regulations by giving three examples, and the identification of ethical responsibilities in blockchain.

\item  We propose a blockchain governance framework {that} categorises the %adopted
governance structures according to {the decentralisation level of governed blockchains}. The framework extensively covers the existing governance structures. 

\item The framework provides a systematic mapping of blockchain stakeholders and their respective decision rights, accountability, and incentives. This mapping can help inform stakeholders and {the} broader public about the authority, capability, and responsibility in blockchain governance.

\item The framework extends governance to the blockchain ecosystem, which is comprised of four layers: data, platform, application, and community. We identify a set of governance mechanisms and associate each of them with {the} respective stage of each layer's lifecycle.

\item The framework emphasises legal compliance and ethical responsibilities in blockchain governance. Using practical examples, we {analyse the integration of existing regulations with the framework,} and {discuss} the ethical responsibilities {in} blockchain governance.

\end{itemize}

We perform case studies on five well-known blockchain platforms: Bitcoin, Ethereum, Dash, Tezos, and Hyperledger Fabric. We apply our framework to these platforms to analyse how they implement governance, and examine whether our principles are considered in {a} real-world context. We distinguish similarities and differences across these platforms and identify current gaps. The results show that our framework is feasible and applicable. In addition, we compare the proposed framework with nine existing blockchain governance frameworks.

%We evaluate the proposed framework using {5} well-known blockchain platforms: Bitcoin, Ethereum, Dash, Tezos, {and Hyperledger Fabric}. 

%The main contributions of this paper are as follows: 1) a blockchain governance framework consisting of six principles which can provide holistic guidance to practitioners on important governance issues for the development of governance-driven blockchain systems; 2) analysis on the allocation of decision rights, accountability, and incentives in regard to the involving stakeholders, to identify their authority, capability, and responsibility; 3) extension of governance to the overall blockchain ecosystem, which is comprised of four layers (i.e., data, platform, application, and community). The proposed framework is evaluated via case studies to validate the feasibility and applicability.

% \begin{itemize}
%     \item We design a blockchain governance framework which can provide holistic guidance to practitioners on important governance issues for the development of governance-driven blockchain systems.
    
%     \item We identify six core governance principles with consideration of blockchain's inherent characteristics.
    
%     \item We analyse the allocation of decision rights, accountability, and incentives in regard to the involving stakeholders, to determine their authority, capability, and responsibility.
    
%     \item We provide high-level governance design which is extended to the ecosystem of blockchain throughout the full lifecycle.
% \end{itemize}

The remainder of this paper is organised as follows. The next section provides a literature review. Section~\ref{sec:methodology} explains our research methodology. Section \ref{sec:framework} presents our blockchain governance framework. Section \ref{sec:evaluation} demonstrates {the} qualitative analysis of {our proposed} framework. Section \ref{sec:conclusion} concludes the paper and outlines future work.

%\section{Preliminaries}
%\label{sec:preliminaries}

%\subsection{Blockchain Technology}
%In recent years, blockchain has become a hot topic to develop next generation applications in a decentralised way, acting as both a distributed ledger and a computing infrastructure.

%Being a distributed ledger in software architecture design, a blockchain network can verify and store transactions~\cite{scheuermann2015iacr}. Without relying on any central trusted authority, all participants in a blockchain network need to reach consensus on transactional data states to achieve trust. In the bitcoin consensus mechanism proposed by Nakamoto, the majority of nodes should be honest to against third party intermediaries through game theoretic incentives~\cite{Satoshi:bitcoin}. A blockchain is build up by a list of identifiable blocks linking to the previous one in a chronological order. Each block contains transactions which record the changing states of data. 

%Smart contract enables on-chain computing, extending the capability of blockchain technology. Smart contracts~\cite{Omohundro:2014} are programs deployed and running on blockchain, which can express triggers, conditions and business logic to support more complex programmable transactions.

%\subsection{Blockchain Governance}

%existing governance frameworks and how they differentiate blockchain governance.

\section{Related Work}
\label{sec:relatedwork}

%We performed a systematic literature review to understand the state-of-the-art of blockchain governance~\cite{liu2021systematic}. Meanwhile, we also reviewed existing governance frameworks and standards. 

We reviewed existing governance frameworks and standards, including IT governance~\cite{weill2004governance, cobit2012business}, data governance~\cite{ballard2014ibm, ISO38505}, {open-source software} (OSS) governance~\cite{o2007emergence, de2007governance}, platform ecosystem governance~\cite{tiwana2010platform}, and corporate governance~\cite{visa}. IT governance focuses on how to align information technology with organisational goals in enterprises~\cite{weill2004governance}. Data governance narrows down the subject of governance to data, specifically, how organisations evaluate, direct, and monitor the use of data~\cite{ISO38505}. OSS governance analyses how a group of volunteers contribute to the peer production of software~\cite{de2007governance}. Platform ecosystem governance provides {platform-agnostic} guidance on how decisions are made about a platform {regarding} environmental dynamics~\cite{tiwana2010platform}. Corporate governance highlights code of conduct to retain the trust of clients~\cite{visa}.

We found that the decentralisation nature of blockchain differentiates its governance from existing governance frameworks with the absence of a clear source of authority. For instance, in both IT governance and corporate governance, there is a clear governance structure describing the allocation of responsibility, capability, and authority within an organisation. Policies and strategies are always decided by top executives, which embodies the centralisation of power. {Further,} governance {is realised} for the benefit of this particular organisation. Data, open-source software (OSS), and platform ecosystem governance have a similar situation {in} that they all {involve} multiple organisations, and there is a trusted entity acting as the agent between {all relevant} organisations. 

Compared to the above frameworks, blockchain itself has intrinsic characteristics, such as consensus, incentives, transparency, and immutability, which could potentially create a situation where they conflict with human values (e.g. privacy, security), necessitating a trade-off analysis when implementing governance. In general, the decentralised environment gives more power to {the} community and highlights the notion of democracy. All on-chain transactions are under the algorithmic governance predefined by {the} blockchain project team, while other stakeholders can shift the direction of a blockchain via voting. Since there {are} no trusted third parties to coordinate the business relationships, the governance process may involve complex coordination among differential stakeholders. Consequently, reaching consensus from diverse intentions is considered the key point of blockchain governance. Besides, the codebase of blockchain is relatively static due to the immutable data structure design. Based on {the} above factors, any revision to {the blockchain platform code} or on-chain historical data will lead to a voting process for proposal acceptance and hard fork to update the on-chain protocol. Accordingly, the concepts of self-governance~\cite{ostrom2009understanding} and adaptive governance~\cite{olsson2006shooting} are consistent with blockchain governance to satisfy the needs of decentralisation and inherent democracy spirit, and desires for risk management.

\begin{figure*}[t]
	\centering
	\includegraphics[width=0.75\textwidth]{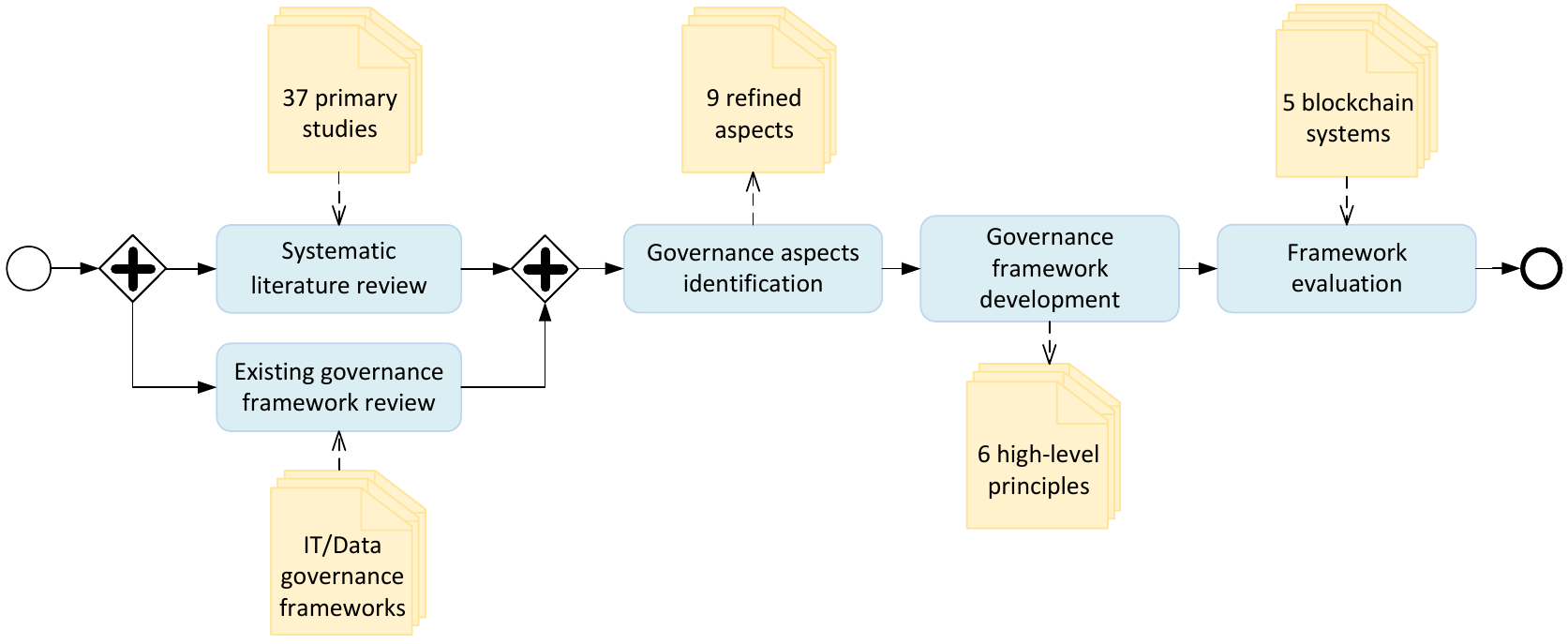}	
	\caption{Overview of the research process.}
	\label{fig:methodology}
\end{figure*}

Further, there are nine studies that proposed governance frameworks for blockchain. Katina et al.~\cite{selected5} propose seven interrelated elements (philosophy, theory, axiology, methodology, axiomatic, method and applications) for further exploration of this topic. Allen and Berg~\cite{selected7} provide a descriptive framework to understand the exogenous and endogenous governance in blockchain. John and Pam~\cite{selected10} and Pelt et al.~\cite{selected11} both study the on-chain and off-chain development processes to realise governance. Beck et al.~\cite{selected14} formulate a blockchain governance framework that is centred {on} the three dimensions of decision rights, accountability, and incentives adopted from IT governance. Howell et al.~\cite{selected15} focus on the membership and transacting relationships. Werner et al.~\cite{selected31} develop a taxonomy of platform governance for blockchain. Hofman et al.~\cite{hofman2021blockchain} propose a high-level analytic framework, regarding six aspects (i.e. why, who, when, what, where, how) in the governance of blockchain. Tan et al.~\cite{TAN2021101625} analyse this research topic from the perspective of social sciences.

Nevertheless, existing frameworks do not present an extensive view of blockchain governance. Consequently, they can only provide limited guidance to the community with scattered governance mechanisms. In this paper, we propose a comprehensive blockchain governance framework, which formulates nine core governance aspects into six high-level principles, to support a better governance process for the overall blockchain ecosystem.

\section{Methodology}
\label{sec:methodology}

%We employ the grounded theory~\cite{groundedTheory} in this study, which can provide guidance on absorbing and delivering information from real-world practices on a certain topic~\cite{Gregory2011}. Specifically, we adopt the qualitative methods in empirical studies of software engineering \cite{799955}.  Fig.~\ref{fig:methodology} illustrates the steps of our methodology, including review on related literature and existing governance frameworks and standards, blockchain governance framework design, and case study.

This section introduces the methodology of this study. As shown in Fig.~\ref{fig:methodology}, this study is conducted through mainly five steps. First, a systematic literature review (SLR) was performed following Kitchenham's guidelines \cite{keele2007guidelines} to understand state-of-the-art of blockchain governance. Secondly, we also reviewed extant governance frameworks and standards, to analyse the characteristics of blockchain governance. Afterwards, we identified six high-level governance principles, and proposed a new blockchain governance framework. Finally, we performed {a} qualitative analysis to confirm the feasibility and applicability of our proposed framework.

% which was then evaluated by applying it to analyse the governance implementation of {5} well-known blockchain platforms based on their %open 
% websites and documents.

%\subsection{Generation of theory}

The SLR consists of four main steps: $\bullet$ keyword search, $\bullet$ selection based on predefined criteria, $\bullet$ backward and forward snowballing, and $\bullet$ data extraction and synthesis. We collected the literature from five reference databases: ACM Digital Library, IEEEXplore, ScienceDirect, SpringerLink, and Google Scholar. 1061 papers were retrieved, while after the removal of duplicates, and title/abstract screening, there were 75 papers in the initial selection set. We then conducted a full-text screening, and scoped down to 27 papers as the tentative selection. Subsequently, the snowballing process identified 10 more papers, and the final selection set had 37 papers for data extraction, analysis, and synthesis. Table~\ref{tab:SLR} illustrates our inclusion/exclusion criteria. The primary studies were analysed with six research questions (i.e. what, why, where, when, who, and how), which cover core aspects of blockchain governance~\cite{liu2021systematic}. The major findings of this SLR are as follows: 
\begin{itemize}
    \item Governance is applied for blockchain to address software qualities and human values such as adaptability, upgradability, security, {and} fairness. The ultimate goal of preserving these attributes is the prosperity of blockchain in various application scenarios by improving the trustworthiness of a blockchain platform.
    
    \item Blockchain governance can be split up into on-chain and off-chain regarding where governance methods are enforced. Specifically, on-chain governance aims at the operations and decisions for {a} blockchain platform itself and {the} stored data, while off-chain governance emphasises collaborations of {the} blockchain community in {the} real world.
    
    \item The community consists of different stakeholders who join a blockchain platform and hold the rights for governance issues. Major stakeholders include {the} project team, node operators, users, application providers, and regulators.
    
    \item Governance methods can be categorised into process mechanisms and product mechanisms. The former type supports blockchain development by determining meta-rules for governance, whilst the latter ones are coded as features and functionalities of blockchain to regulate the behaviour of stakeholders.
\end{itemize}

\begin{table}[!h]
\footnotesize
\centering
\caption{Selection of Primary Studies in SLR.}
\label{tab:SLR}
\begin{tabular}{p{0.1\columnwidth}p{0.8\columnwidth}}
\toprule

\multirow{3.5}{0.1\columnwidth}{\bf Inclusion criteria} &
A paper that proposes a solution for the governance of blockchain. \\
\cmidrule(l){2-2}

& A paper that proposes principles or frameworks for developing governance of blockchain. \\ 
\cmidrule(l){1-2}

\multirow{9}{0.1\columnwidth}{\bf Exclusion criteria} &
Papers that focus on ``governance through blockchain" instead of ``governance of blockchain". \\
\cmidrule(l){2-2}

& Older version of a study that has a more comprehensive version. \\ 
\cmidrule(l){2-2}

& Papers that are not written in English. \\ 
\cmidrule(l){2-2}

& Papers that are not accessible. \\ 
\cmidrule(l){2-2}

& Survey, review and SLR papers. We do not conduct data extraction or synthesis from these studies as they are considered the related work of this literature review. \\ 

\bottomrule
\end{tabular}
\end{table}

We reviewed extant governance frameworks and specifications, as discussed in Section~\ref{sec:relatedwork}. We included the literature of IT governance, data governance, and platform ecosystem governance adhering to a previous work of data governance for platform ecosystem process management~\cite{sunny_lee}, whilst OSS governance~\cite{selected11, selected20} was selected based on the primary studies in our SLR. For corporate governance, VISA was selected as it is often compared to blockchain platforms regarding financial issues.

We identified nine core aspects of blockchain governance from the literature review, including decentralisation level, incentive, decision rights, accountability, stakeholder, ecosystem, lifecycle, legal compliance, and ethical responsibility. The identified aspects are then formulated into a blockchain governance framework, consisting of six high-level principles, as follows.%, and a framework {established} by further investigating and integrate refined factors within these principles. 

\begin{itemize}
    \item Principle 1 elucidates different decentralisation {levels} of blockchains, which is extracted from the research question \textit{``What is blockchain governance?"} in our SLR. {The} decentralisation level determines both the underlying infrastructure and governance structure of a blockchain.
    
    \item Principles 2-4 explore the mapping of incentive, decision rights, accountability and divergent stakeholders. The former three governance dimensions are extracted from the research question \textit{``What is blockchain governance?"}, while blockchain stakeholders are identified in \textit{``Who is involved in blockchain governance"}).%, to analyse and refine the motivations, authorities, capabilities and responsibilities of involved entities in blockchain governance.}
    
    \item Principle 5 extends governance to the overall blockchain ecosystem with elaborate governance mechanisms and their expected effects. This principle integrates four research questions from the SLR: \textit{``Where is blockchain governance enforced?"} \textit{``When is blockchain governance applied?"} \textit{``How is blockchain governance designed?"} \textit{``Why is blockchain governance adopted?"}
    
    \item Principle 6 is extracted from the research question \textit{``Why is blockchain governance adopted?"} Legal and ethical responsibilities represent the essential guidelines for human behaviours in {the blockchain} ecosystem.
\end{itemize}

We demonstrate {a} qualitative analysis of the proposed framework. We apply the framework to five blockchain platforms as case studies, to confirm its feasibility and applicability. {The} selected blockchain platforms are Bitcoin, Ethereum, Dash, Tezos and Hyperledger Fabric. Bitcoin and Ethereum are included regarding their market values and active users, while Dash and Tezos blockchain are chosen because they {have} representative novel governance structures. Specifically, Dash introduces the concept of ``masternodes", and Tezos enables smooth on-chain protocol replacement instead of hard fork. Hyperledger Fabric is a representative case of permissioned public blockchains. Data collection was performed on the official websites and documents of these five blockchains. We analysed the collected data using our framework design, to scrutinise the implementation of governance in different blockchain platforms. Afterwards, we compare nine existing blockchain governance frameworks with our framework, which are retrieved from our previous SLR~\cite{liu2021systematic} and continuing literature review. We use identified governance aspects as the comparison factors. %, which can also examine the feasibility and applicability of our proposed framework.

\begin{figure*}[!h]
	\centering
	\includegraphics[width=0.98\textwidth]{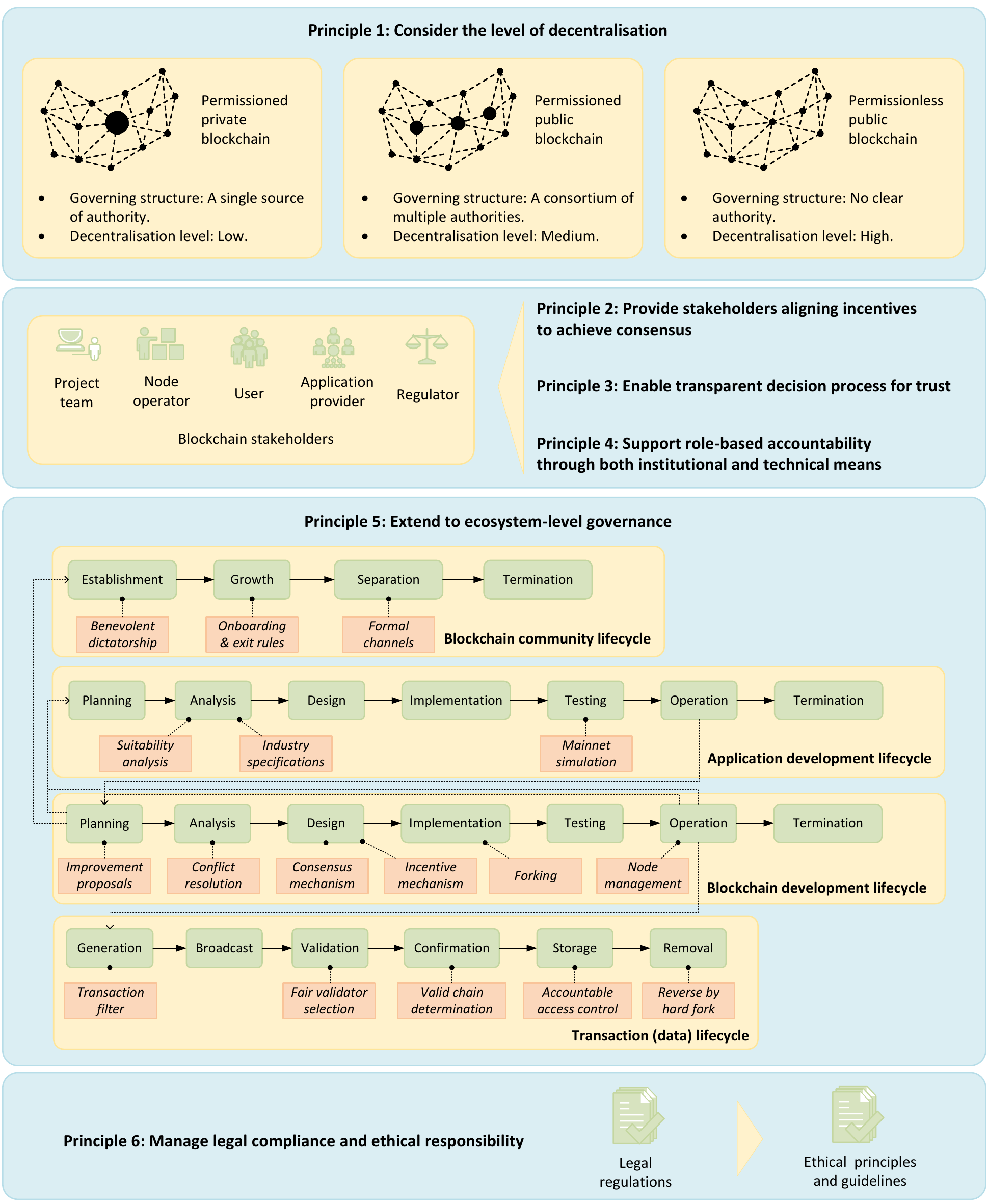}
	\caption{Blockchain governance framework.}
	\label{fig:framework}
\end{figure*}

\section{Blockchain Governance Framework}
\label{sec:framework}

In this section, we present a framework to provide guidance for all relevant stakeholders, to support a better governance process, especially for practitioners as the discussion involves technical aspects of blockchain-based architectures and applications. The overview of our blockchain governance framework is illustrated in Fig.~\ref{fig:framework}, which consists of six principles. %Each principle covers a certain perspective in blockchain and jointly forms the comprehensive governance framework.

%\subsection{Principle 1: Enable transparency and decentralisation}
\subsection{Principle 1: Consider the level of decentralisation}

%Blockchain taxonomy: public/private, permissioned/permissionless

%different level of decentralisation, data transparency

%Blockchain is a broad concept which realises distributed ledger technology in the form of a chain of blocks. 
Blockchain platforms can be classified into three types to meet different requirements. Different types of blockchain may have {a} divergent focus on certain attributes, e.g., performance, flexibility. These types also reflect the different levels of decentralisation, and affect the governance structure regarding allocation of decision rights, accountability, and incentives. The essential classification criteria {include} permission of joining a blockchain, and {the} extent to which users are allowed to participate {in} the operation of a blockchain. They both lead to differences among three blockchain types, such as cost efficiency, performance, flexibility, etc. Hereby, ``operation" denotes services a blockchain provides (e.g. generating and reading transactions), and contributions to the use of blockchain (e.g. the capability of installing a full node).

% \begin{itemize}
%     \item \textit{Permissioned Private Blockchain}
% \end{itemize}

\textbf{Permissioned private blockchain.}
%We start with the introduction of permissioned private blockchain, as it 
Permissioned private blockchains have the lowest level of decentralisation among the three types. The access and operation of a permissioned private blockchain are both restricted to certain entities. This type of blockchain platform has an explicit single source of authority, and is deployed within an institute/organisation. Hence, there is a clear hierarchy structure between stakeholders in blockchain, which should align with the off-chain corporate structure. All involved individuals are identifiable and have fixed user privileges. Compared to the other two types, governance of permissioned private blockchains is more flexible and requires less effort to orchestrate decisions from different stakeholders, as every choice directing the blockchain is made by entities on the top of interior hierarchy.

% \begin{itemize}
%     \item \textit{Permissioned Public Blockchain}
% \end{itemize}

\textbf{Permissioned public blockchain.} 
Permissioned public blockchains only require approval for participation, while operation-related rights are open to the participants. The applicable business context of this type of blockchain is usually a consortium of several organisations maintaining a cooperative relationship. Identity verification in permissioned public blockchains is similar to permissioned private ones. The governance structure is comprised of the hybrid of multiple authorities, each of which may make decisions for its own organisational benefits. Consequently, governance in this type of blockchain usually focuses on the coordination of involved authorities for collective goals.

% \begin{itemize}
%     \item \textit{Permissionless Public Blockchain}
% \end{itemize}

\textbf{Permissionless public blockchain.}
In permissionless public blockchains, the entries and operations are both open to entities, as there {are} no preset privileged stakeholders. Such openness, however, may result in higher complexity of governance in terms of two main aspects: 1) Since each individual is recognised via on-chain pseudonymous addresses without real-world identification, stakeholders may be unwilling to bear the responsibility of making decisions; 2) Stakeholders may need to go through a cumbersome negotiation process to reach {an} agreement, as everyone focuses more on personal interests, which may cause conflicts in decision-making processes. Consequently, permissionless public blockchains usually improve the governance participation rate by incentives, and exploit consensus mechanisms for decision finalisation.

\begin{table*}[h!]
\footnotesize
\centering
\caption{Distribution of 3 governance dimensions over blockchain stakeholders for Principles 2-4.}
\label{tab:incentive}
\begin{tabular}{p{0.09\textwidth}p{0.26\textwidth}p{0.27\textwidth}p{0.18\textwidth}}
\toprule
\bf{Stakeholders} &
\bf{Decision rights} &
\bf{Accountability} &
\bf{Incentives} \\
\midrule

Project team &
%& Financial support; \par
%Architecture design: \par
Platform development: \par
\quad Blockchain infrastructure setting; \par %(block size and interval, smart contract, API); \par
\quad Consensus mechanism; \par
\quad Incentive mechanism;\par
%\quad Identity management;\par
%\quad API design;\par
% \quad Voting mechanism; \par
% \quad Cryptography techniques; \par
% \quad Synchronisation rules; \par
% \quad Participation \& exit rules; \par
% \quad Key management; \par
% \quad Communication channel; \par
%Code implementation; \par
%Platform maintenance: \par 
Project management: \par
\quad Conflict resolution rules; \par
\quad Formal communication channel; \par
\quad Onboarding \& exit rules; \par
\quad Risk management. \par
%\quad Platform monitoring; \par
%\quad Improvement proposal approval; \par
%\quad Forking implementation.
&
Institutional means: \par
\quad Real-world identity verification; \par
\quad Traceable code contributors; \par
\quad Standardised documentation; \par
Technical means: \par
\quad Address-based on-chain identity; \par
\quad Ledger-enabled operation logs.
& Increase of market values; \par
Block rewards; \par
Transaction fees; \par
Service fees.
\\
\cmidrule{1-4}

Node \par operator 
& Replica storage; \par
Block validation; \par 
%Transaction inclusion \& block validation; \par  
Improvement proposal voting; \par
Forking (instance installation). &
Technical means: \par
\quad Address-based on-chain identity; \par
\quad Ledger-enabled operation logs; \par
Institutional means: \par
\quad Real-world identity verification.
& %Rewards set by incentive mechanism
Block rewards; \par
Transaction fees.
\\
\cmidrule{1-4}

User 
& Transaction submission; \par 
Improvement proposal voting. 
&
Technical means: \par
\quad Address-based on-chain identity; \par
\quad Ledger-enabled operation logs; \par
Institutional means: \par
\quad Real-world identity verification.
& Achievement of personal goals: \par
\quad Personal investment; \par
\quad On-chain trading; \par
\quad Data storage.
\\
\cmidrule{1-4}

Application provider 
%& Achievement of business targets; \par Increase on the market value
& Blockchain adoption; \par
%Application monitoring; \par
%Risk assessment \& mitigation: \par
%\quad Application monitoring; \par
Improvement proposal voting. \par
Onboarding \& exit rules.
&
Institutional means: \par
\quad Real-world identity verification; \par
\quad Commercial agreement.
%Technical means: \par
%\quad Address-based on-chain identity.
& Increase of market values; \par
Service fees.
\\
\cmidrule{1-4}

Regulator 
& Risk assessment \& measurement: \par

\quad Regulatory policy; \par %(juristical perspective); \par 
%\quad Law enforcement; \par
%\quad Software license; \par
\quad Audit trail.
 & %\par
%Platform monitoring &
Institutional means: \par
\quad Legal regulation; \par
\quad Real-world identity verification; \par
Technical means: \par
\quad Address-based on-chain identity; \par
\quad Ledger-enabled operation logs.
& Taxes and fees.
\\

\bottomrule
\end{tabular}
\end{table*}

\subsection{Principle 2: Provide stakeholders aligning incentives to achieve consensus}

%Stakeholder goals and incentives
%(figure or table)

%incentives are govern to encourage the participation for governance. upstanders instead of bystanders.

%Originally, incentives were distributed to the block validators (i.e. miners) 

%The emergence of blockchain was along with the triumph of cryptocurrencies (e.g., bitcoin, ether). In cryptocurrencies, incentives refer to distribution of tokens to participants who follow preset mechanisms and contribute to validation of blocks. 
The second principle is to be addressed to a wide range of blockchain stakeholders. There are collective areas aligning diverse stakeholder objectives, giving incentives wide applicability. In blockchain governance, incentives are generalised to factors that may influence stakeholders’ behaviours. The governance structure needs to provide incentive mechanisms to drive desirable behaviours (e.g., participation {in} governance), and resolve conflicts between stakeholders (e.g., finalising a decision), which usually end up with the finding of a Schelling point. In permissionless public blockchains, incentive mechanisms may incorporate game theory to align different directions of blockchain governance. In this case, incentives are mostly rewarded on-chain, hence, can be traced by any other stakeholder. Meanwhile, incentives may be distributed off-chain according to institutional rules preset by {the} top executives in permissioned private blockchains, or based on commercial agreements among {the} multiple authorities in permissioned public blockchains.

We conclude {the} incentives of different stakeholders in Table~\ref{tab:incentive}. In general, monetary incentives are straightforward to reward stakeholders in governance-related issues. {The} project team and application providers can benefit from {the} increase {of} market values (e.g., increase of cryptocurrency price), and service fees. 
Node operators receive block rewards and transaction fees according to predefined incentive mechanisms for their contributions {to} blockchain operation, while {the} project team may also obtain a certain proportion as compensation. Incentives of blockchain users mainly come from their intentions to use blockchain and applications, including personal investment, on-chain trading, and data storage. Finally, regulators collect taxes and charge for regulation services (e.g., audit).
% enforce regulations or are paid for certain services such as auditing. %Finally, the incentive of blockchain users is not apparent as it relates to the usage scenario of a blockchain. In the case of cryptocurrencies, the incentive is also reflected in the increase to token price.

\subsection{Principle 3: Enable transparent decision process for trust}

%describe the decision rights of stakeholders
The decentralisation nature of blockchain drives the need {for} collective decisions on governance issues. %Making a decision 
Decision-makings rely on {the} authorities, responsibilities, and capabilities of relevant stakeholders. Notice that a stakeholder's role may change throughout the lifecycle of a blockchain, which implies the transition of decision rights. We summarised {the} decision rights of blockchain stakeholders in Table~\ref{tab:incentive}.

%The decision rights of %different 
%blockchain stakeholders are summarised in Table~\ref{tab:incentive}.

The project team of a blockchain mainly comprises developers and the foundation, for technical and monetary support respectively. Most governance-related decisions for a blockchain platform are made by developers. Specifically, developers' decisions determine a series of on-chain and off-chain meta-rules for both blockchain platforms and communities. On-chain decisions include {the} setting of blockchain infrastructure (e.g., block size and interval), and design of consensus and incentive mechanisms, which can regulate the behaviours of other stakeholders when using blockchain. Off-chain decisions refer to {the} implementation of code, establishment of conflict resolution rules, formal communication channel, and {onboarding} and exit rules for the whole community. In addition, blockchain governance should consider risks such as software bugs in coding, cyber attacks in operation, breach of commercial contracts, etc. Risk management in blockchain is achieved by platform update and evolution, which {the} involves approval of improvement proposals, and implementation of new features, also known as forking.

%. Developers need to conduct architecture design, code development and maintenance, determine conflict resolution rules, and handle risk assessment and measurement. Particularly, architecture design includes the setting of block size and interval, consensus and incentive mechanisms, on-chain identity management, and API development. The code maintenance of a blockchain platform is special due to the immutability of a deployed blockchain, hence, the update of source code usually requires forking to enable the new functionalities. The conflict resolution is implemented via the voting process happened both on-/off-chain. In addition, the project team needs to assess the risks in the use of blockchain platform considering the possible attacks and apply measures to mitigate the risks (e.g., real-time monitoring, improvement proposal for platform upgrade). Finally, the onboarding and exit rules of a blockchain platform are also decided by the project team.

%determine and realise the rules within the blockchain platform, including block size and interval, smart contract, API, etc. Particularly, the consensus and incentive mechanisms need to be exquisitely designed as they are the core elements for blockchain governance. When implementing the platform, developers are responsible to build, release, and test the code, while also need to decide how the code is maintained. Further, the project team can finalise the decision of forking.

Node operators hold either local full nodes to maintain all historical ledger contents, or light nodes with compressed information like block header. Meanwhile, an operator can become block validators, or special node operators who are granted rights to make improvement proposals for more contributions. As nodes are the fundamental elements to construct a blockchain, operators are the key roles to finalise forking by choosing and installing blockchain instances of a specific version.

{Users interact} with blockchain services via submitting transactions to achieve personal goals. When encountering an abnormal function/service, a user can report to {the} project team for improvements, while other users can then join the voting of proposed solutions. 

Application providers have the right to choose which blockchain platform to be adopted in their existing workflow regarding business targets. They can contact {the} project team to exchange domain knowledge and requirements for blockchain adaptation, which may be raised in the form of {an} improvement proposal. Further, an application may have its own setting for onboarding and exit rules, which should be decided by providers.

%They can contact the project team periodically to exchange domain knowledge for the adaptation of blockchain. The risk assessment and measurement of the application involves the evaluation to internal blockchain component.

%As they monitor the overall performance of applications, the blockchain component is also supervised. 

Regulators mainly refer to governments and third-party auditors, who are responsible for ensuring the compliance of laws and ethics. The assessment and mitigation of corresponding risks rely on {the} enactment of regulatory policies and audit {trails} in blockchain ecosystem.

%Governments can issue policies as the meta-rules (e.g. GDPR to protect data privacy), and software license for the development and release of a blockchain platform. Auditors set the policies related to conduct and report the auditing activities of blockchain. Blockchain itself saves the operation log at the full nodes which can be used for auditing activities. These regulators have the right to ensure the risk mitigation for compliance of laws and ethics.

%These regulators have the right to monitor the blockchain platforms to ensure they comply with the corresponding laws and further, ethical rules.

Ensuring the transparency of decision-making processes is critical in blockchain governance for stakeholders to {oversee} whether decisions are reasonable, accordingly, to gain trust from {the} whole community. When decisions are made on-chain, blockchain itself can record individuals' choices and the eventual outcome. This can usually be found in {permissionless} blockchains, where decision rights are allocated to different stakeholders for fairness and democracy in a highly decentralised environment. Off-chain decision-makings should be conducted in formal channels which are open to relevant stakeholders. Note that the clear source of authority may integrate multiple roles and corresponding decision rights into certain individuals in {a} permissioned context where significant decisions are usually made by blockchain deployers.

\subsection{Principle 4: Establish role-based accountability through both institutional and technical means}

Accountability in blockchain governance refers to the identifiability and answerability of stakeholders for their decisions. Accountability should be enforced in an explicit way to manifest the allocation of stakeholders' responsibilities and capabilities. Failure on this governance dimension may result in the abuse of power and eventually, {the collapse of the entire blockchain system} due to the risk of centralisation.

Accountability for blockchain governance can be established via both institutional and technical manners. Institutional accountability is applicable to conventional governance structures in a blockchain, which requires standardised documentation to report {the} off-chain accountability process. Technical implementation of accountability is embedded in the distributed ledger, providing evidence for on-chain accountability by recording both data and operations. In comparison with off-chain accountability, the on-chain process is still less mature and relies on how developers design the blockchain platform to reduce risks and uncertainties. 

Accountability is also dependent on the decentralisation level of a blockchain. In permissioned blockchains, it is compulsory to conduct identity verification for entities to join the blockchain. Hence, it does not cost {much effort} to recognise accountable individuals for certain events or operations. By contrast, participants identify themselves via pseudonymous accounts in permissionless blockchains, which may increase the uncertainties of finding responsible entities and hence, hinder the accountability process.

As shown in Table~\ref{tab:incentive}, the accountability process of {the} project team, application provider, and regulator relies more on institutional means, while technical means are more important for node operator and user as they are usually identified via on-chain addresses. In general, institutional means of accountability enables {an} effective and efficient decision-making process where the source of authority is clear. Relevant stakeholders should be considerate when making governance decisions, otherwise, they could be criticised and punished immediately for their neglect of duty, which may further affect real-world business or individual reputation. For different stakeholders, institutional means may vary. For instance, the accountability of {the} project team relates to institutional means regarding the visibility and accessibility of source code development to {monitor} contributors. While for application providers, institutional means are dependent on their organisational code of conduct. However, technical means of accountability do not ensure such {a} rigorous decision-making process as the cost of violation may be low. A malicious node operator may attempt to forge and broadcast invalid blocks to the network. This can be traced by on-chain ledger logs {to impose a fitting penalty} such as blacklisting the blockchain account. Nevertheless, the accountable entity can make more attacks over and over again via other blockchain accounts.

\subsection{Principle 5: Support ecosystem-level governance}

The governance of blockchain does not locate in merely the platform itself, but needs to support a broader environment of {the} overall ecosystem. Platform is the base of blockchain ecosystem, which includes underlying on-chain data, {the} superstructure of blockchain-based application systems, and the whole community comprising of different stakeholders. We illustrate {the} main phases in the lifecycle of each layer in Fig.~\ref{fig:framework}.

%In each layer, we list high-level questions as the guide to evaluate governance-related issues in blockchain ecosystem.

% \begin{itemize}
%     \item \textit{Platform Layer}
% \end{itemize}

\textbf{Platform layer.}
%\subsubsection{Platform layer}
Platform is both the base and target of governance structure for blockchain ecosystem. All governance-related design and tactics will eventually take effect in this layer. We adopt the software development process~\cite{softwaredevelopmentprocess} to describe different phases of this layer.

During the initial planning and analysis stages of a blockchain platform, business context is taken {into} consideration to determine the blockchain type. Later in the design phase, the blockchain architecture with all on-chain self-governance features is constructed (e.g., consensus and incentive mechanisms), along with {the} distribution of stakeholders' decision rights, accountability, and incentives. When developing the platform, management of source code should be decided (e.g., open or closed source), while the testing environment needs to be separated from the mainnet to avoid influencing {the} actual business process. When a blockchain platform is officially in use, all on-chain issues need to adhere {to} established mechanisms of previous phases. Note that the management of nodes is specified in Principles 2-4 from the perspective of node operators.

%In previous development process, the project team may act as dictator when there is no other stakeholders. When a blockchain platform is officially in use, the community becomes larger with the participation of differential stakeholders, who will join the consequent decision-making process. 

If a blockchain platform needs updates or adjustments within a risk management process, the governance structure in permissioned blockchains can make quick responses such as forking or protocol replacement. In {the} case of permissionless blockchains, an improvement proposal is created, which starts a new epoch of development process from {the} planning phase. Such upgrades and evolutions require a series of formalised development procedures. An improvement proposal should be analysed by eligible stakeholders, with proper conflict resolution measures, usually voting, for its acceptance or rejection. The accepted ones are then codified and integrated into source code, and released as new versions via forking. This iteration of blockchain upgrades embodies an adaptive governance process to manage and mitigate risks in blockchain operation. Finally, blockchain {platforms} may be terminated for certain reasons. At this point, blockchain governance focuses on coordination among stakeholders about the transfer or redistribution of on-chain assets and resources, which requires explicit {descriptions} and {explanations} of decisions to guide the whole process. 

Additionally, blockchain governance structures should define inter-system interoperation policies between different platforms regarding the decentralisation level. This will reflect in {the} data flow, integration of business process, and even adjustment of decision rights process. For instance, if two permissionless blockchain platforms have interoperations, the extent of visible and shared distributed ledger contents, transition of transaction format, {and} compatibility of on-chain programmability (i.e. smart contracts) should all be scrutinised.

%(change a name: enable on-chain and off-chain governance)

%platform: consensus protocol, infrastructure
%design (block size, interval), token or not (miner), cross-chain %interoperation blockchain lifecycle

% \begin{itemize}
%     \item \textit{Data Layer}
% \end{itemize}

\textbf{Data layer.} 
%\subsubsection{Data layer}
%Conventionally, the governance of data involves 5 phases as shown in Fig.~\ref{fig:ecosystem} \cite{ISO38505}. While 
In blockchain, data governance is carried out along with {the} transaction lifecycle. When designing a blockchain platform, on-chain data governance is covered in {the} design phase, for instance, {a} standardised format to ensure transaction quality, access {policies} for security and privacy, and sharding technique for scalability. During the operation of a blockchain platform, transactions are generated and broadcast when users send data to blockchain. Insofar, a filter is utilised to automatically discard transactions that do not meet the unified requirements (e.g., transaction format, data contents). Transactions are then kept in {nodes' local memory} pool waiting for validation. After being validated and confirmed, data is then officially stored in blockchain. Hereby, confirmation can resolve conflicts that there may be several blocks generated at the same time, which requires subsequent blocks to confirm the workload of validation. The longest chain is viewed as the valid one. Note that pending transactions may be outdated and then automatically removed from the pool.

The emphasis {on} data governance during blockchain operation is placed {on} accountable access control of transactions. Capabilities of writing/reading data to/from blockchain, as well as validating transactions, are assigned to specific stakeholders considering {the} deployed blockchain type. For instance, in permissioned blockchains, validators are often appointed by authorities, while novel consensus {mechanisms are} designed and implemented to impartially elect the validator each round in permissionless blockchains. Changes of data states are logged by blockchain transactions, which enables {an} on-chain accountability process regarding all decentralisation {levels}. However, reading transparent data usually leaves no trace to blockchain history as no transaction is sent, which may hinder the identification of individuals. 
All on-chain operations should respect the rights of data subjects that illegal data need to be removed from blockchain by conducting a hard fork and {reversing} historical transactions. Finally, when a blockchain is terminated, all on-chain data will be destroyed under predefined guidance.

% \begin{itemize}
%     \item \textit{Application Layer}
% \end{itemize}

\textbf{Application layer.} 
%\subsubsection{Application layer}
Blockchain governance also covers the utilisation of blockchain as a component in software application systems. A blockchain-based application may have {a} close connection to the blockchain platform itself and consequently, its development process almost synchronises with {the} blockchain development process. This situation is usually found in blockchain-based decentralised finance (DeFi). Another situation is that when a blockchain platform is in operation, further investment is made to build up DApps over the platform. Application providers need to conduct suitability analysis regarding domain knowledge, and check whether adopting blockchain is compliant {with} industry standards and regulations during initial procedures. Afterwards, implementation and testing rely on the APIs and testnet provided by {the} blockchain project team. When operating the application, changes to industry regulations and specifications may generate new requirements. which in turn {leads} to upgrades of {the} underlying blockchain platform. Finally, an application may fail, and then needs to remove its data from blockchain and transfer resources based on the instructions of {the} project team or previous contractual agreements.

%which includes two main types of stakeholders: project team and application provider. The project team should developed API for a blockchain to interact with other software components, while an application provider should ensure that the adopted blockchain platform comply with related rules and specifications of the industry, which may involve monitoring the real-time status of blockchain. The adoption of blockchain also requires consideration on background knowledge of a specific application domain and level of decentralisation, for the adaptability and upgradability regarding the varying demands. In the termination stage of a blockchain, the related applications should remove their data and transfer resources from the blockchain, according to the instruction of project team, or contractual agreement between them.

%In a broader context, the application layer may involve interactions between two blockchain platforms. This will reflect in the data flow, integration of business activities, and even the transfer of decision-making process. Blockchain governance structures should define inter-system interoperation policies to another blockchain regarding the different level of decentralisation.

% \begin{itemize}
%     \item \textit{Community Layer}
% \end{itemize}

\textbf{Community layer.} 
%\subsubsection{Community layer}
The start of blockchain platform development indicates the establishment of corresponding off-chain community. In {the} early stages, {the} project team is considered the benevolent dictator as there {are} few other stakeholders. {The} project team can persuade them about certain decisions with its expertise. When a blockchain platform is officially in use, the community becomes larger with {the} participation of differential stakeholders, who will join the consequent decision-making processes. Along with the maturity of a blockchain platform, {the} off-chain community is gradually specialised and then divided into different groups {regarding} their roles and decision rights~\cite{williamson1986economic}.

In particular, {the} project team and application {providers} may exist in the form of a company {respectively}. Accordingly, the internal governance of these organisations can refer to existing specifications. Another significant facet is the training of blockchain practitioners, {which informs the} capability development and growth in the future workforce. Ensuring they have adequate expertise, for the design and implementation of blockchain {platforms} and applications to avoid {flaws}, can gain more trust from other stakeholders. Further, blockchain governance highlights off-chain collaboration among different stakeholders. In permissioned blockchains, there is {a} systematic process to communicate in the constitutional hierarchy. By contrast, permissionless blockchains maintain explicit off-chain governance within the community, which implies the need {for} formal communication channels (e.g., maillist, online forum) to transparentise significant decision-makings (e.g., improvement proposal). Finally, termination of community is along with the collapse of blockchain {platforms}, where proper coordination of stakeholders is needed to rearrange off-chain resources and assets.

%as the DLS matures, and especially as it increase its scale of operations, the roles would be expected to gradually separate out (i.e. specialisation, as per Williamson, 1986 emerges).

%As the project team and application provider may exist in the form of company, the internal governance of these organisations can refer to existing specifications. A significant facet is the training of blockchain developers to they have adequate expertise for the design and implementation of a blockchain platform. In addition, blockchain governance highlights the off-chain collaboration among different stakeholders of a blockchain platform. In permissioned blockchain, the governance may be implicit due to the low decentralisation level. The clear source of authority may integrate multiple stakeholder roles and corresponding decision rights, which involves onboarding and exit rules of permissioned blockchain platforms. By contrast, permissionless blockchain  maintains explicit off-chain governance within the community, which implies the need of formal communication channels (e.g., maillist, online forum) to transparentise the significant decision-makings (e.g., improvement proposal).

%Community: improvement proposal, code management (related to application, and business context, open source or not)

%application: related regulation policies, contractual agreement between the application provider and blockchain project team

%data: data / transaction: transaction quality/format, data lifecycle

\subsection{Principle 6: Manage legal compliance and ethical responsibility}

The final principle ensures that all governance-related decisions and processes conform {to} existing legal {regulations} and ethical responsibilities. Law is usually considered to set the minimum standards of human behaviours. Ensuring legal compliance depends on where blockchain platforms and applications are launched and deployed, and how regulators (e.g. governmental institutions) enact policies to safeguard the blockchain development process and shape the whole ecosystem against risks and uncertainties. Examples include the regulation and standards in Australia in terms of digital identity, data provenance, and taxation~\cite{AustralianGovernment}. The European Commission proposed new law on crypto-assets and plan blockchain regulatory sandbox~\cite{EuropeanCommission}. 
In addition to blockchain-specific laws, other general regulations should also be considered, for instance, the General Data Protection Regulation (GDPR) specifies the ``Right to Be Forgotten"~\cite{voigt2017eu} to protect personally identifiable information when managing on-chain data. Another issue is the review of data {contents} before a transaction is officially included in a block, {to} ensure data quality and avoid malicious information on-chain, such as child pornography. If such information is already stored on blockchain, a hard fork is required to reverse the transaction history.

%combating and preventing criminal activities in decentralised finance such as money laundering and blackmail. The General Data Protection Regulation specifies the Right to Be Forgotten~\cite{voigt2017eu} for the protection of personally identifiable information, which should be considered when storing data on-chain.

Ethical responsibilities denote the maximum standards of human behaviours. Promote and encourage ethical principles and guidelines in blockchain governance can help uphold and protect human values, and the whole ecosystem will be more responsible and trustworthy. For instance, blockchain technology is usually criticised for energy-intensive platforms and applications, especially the ones applying {the} Proof of Work (PoW) consensus mechanism. Consequently, how to reduce energy consumption and benefit {the} environment is regarded as a significant governance topic. Other than environmental well-being, ethical responsibilities also involve consideration {of} human values from the perspective of software engineering, such as privacy, transparency, integrity, etc~\cite{HumanValue}. Embedding ethical responsibilities in blockchain governance is hard to guarantee, and requires the awareness of practitioners over such values and overall community culture.

\begin{figure*}[t]
	\centering
	\includegraphics[width=\textwidth]{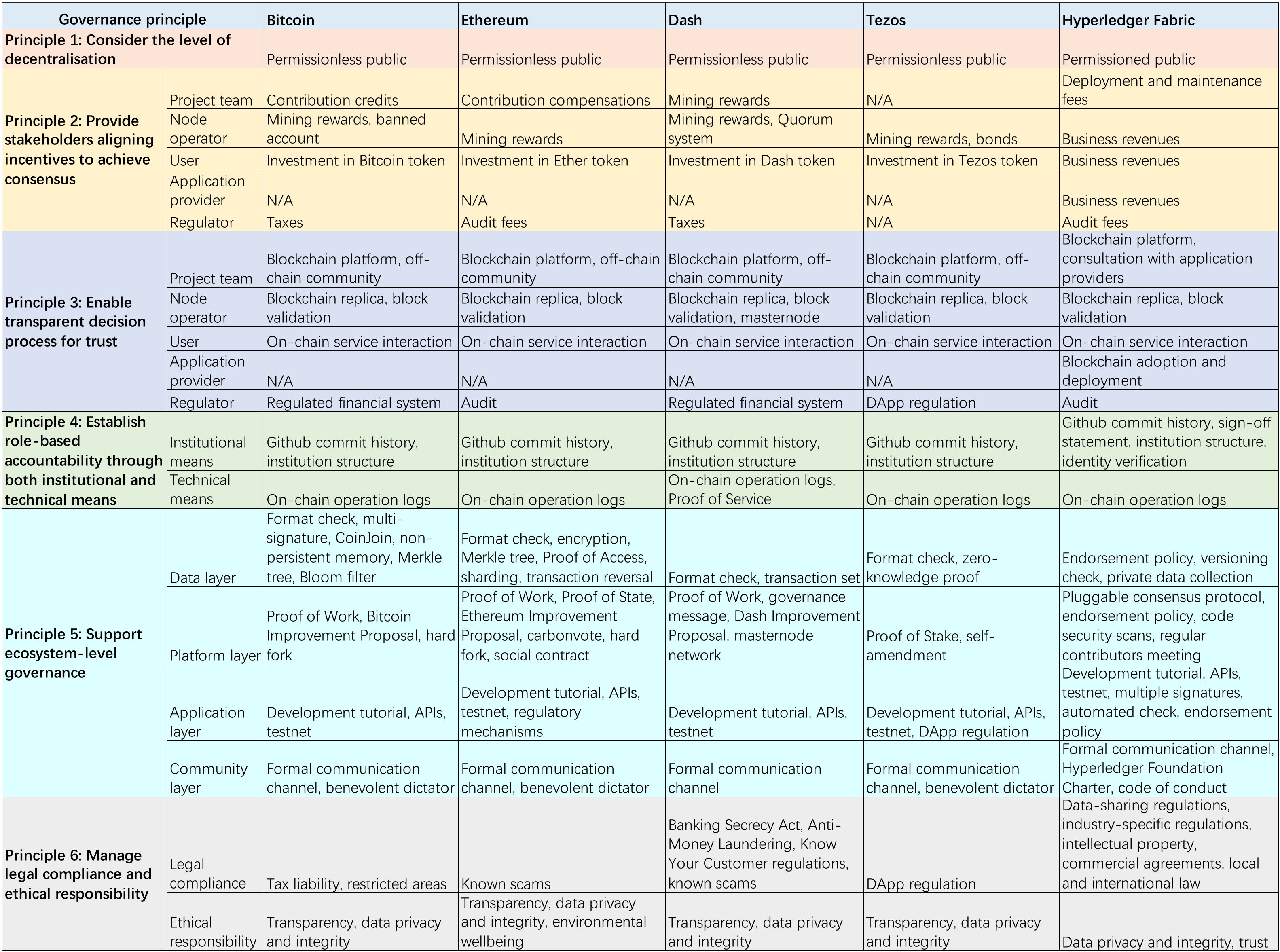}	
	\caption{Case study results. The framework is feasible as all principles are considered in real-world context, and applicable to different blockchain platforms.}
	\label{fig:case_study}
\end{figure*}

\section{Qualitative Analysis}
\label{sec:evaluation}

%In this section, we first present the state of practice of extant blockchain platforms and conduct case studies on them, and then compare our framework with related work. 

In this section, we present {a} qualitative analysis of the proposed framework via case studies on extant blockchain platforms, and comparison with existing blockchain governance frameworks.

\subsection{Case Study}
The purpose of case studies is to understand whether and how blockchain governance is implemented in {a} real-world context, while also {confirming} that our proposed framework is feasible and applicable. Specifically, feasibility refers to whether the principles in our framework solve blockchain governance problems, and applicability implies that the framework can be applied to different blockchain platforms to analyse their governance structures.

We selected five blockchain platforms for case studies: Bitcoin, Ethereum, Dash, Tezos, {and} Hyperledger Fabric. Bitcoin and Ethereum are the two most famous blockchain platforms around the globe, Dash and Tezos are included as they are representative {of} novel governance structures, whilst Hyperledger Fabric is selected as it provides a permissioned blockchain platform for a wide range of industry applications. We collected and analysed their implementation of governance from open official websites and documents (e.g., white paper, yellow paper).
%, through which to evaluate whether the proposed framework is feasible, applicable and usable. %The evaluation criteria is ``sufficiency", to denote whether the governance implementation aligns with the principles, where the possible results are ``implemented" if the documents have elaborate discussion on how to implement the principle, ``partially implemented" if the documents mention certain aspects, and ``not implemented" if related analysis cannot be found. 
The rest of this section illustrates the results of case studies, which are summarised in Fig.~\ref{fig:case_study}.
%Table~\ref{tab:Bitcoin} to~\ref{tab:fabric}.

% of governance principle implementation in these blockchain platforms.

% include ``sufficiency" and ``existence". If the collected data provide information regarding the principles, we conclude the practices satisfy ``existence". We further examined that 

% \begin{itemize}
%     \item \textit{Principle 1: Consider the level of decentralisation}
% \end{itemize}

%\textbf{Define blockchain type and decentralisation level.}
\textbf{Analysis of Principle 1.}
For the first principle, Bitcoin, Ethereum, Dash and Tezos are all permissionless public {blockchains}, and have {a} close connection to the DeFi field. In their white papers, {the} project team is the primary governing role to make decisions, whilst other stakeholders are also involved in governance issues, especially node operators. Hyperledger Fabric offers permissioned public blockchain solutions in enterprise contexts. Its project team only provides technical support to the underlying blockchain technology, while organisations that deploy Hyperledger Fabric instances are the authorities for governance-related decision-makings.

%Hence, they are all marked as ``implemented" for this principle.

%Hence, the governance structure involves diverse stakeholders within the blockchain community.

% \begin{itemize}
%     \item \textit{Principle 2: Provide stakeholders aligning incentives to achieve consensus}
% \end{itemize}

%\textbf{Distribute incentives to drive governance.}
\textbf{Analysis of Principle 2.}
In terms of providing incentives, the four permissionless platforms mainly discuss {the} investment income of users, and rewards to both node operators for their contributions {to participate in} the consensus mechanism, and {the} project team as funds for future development. However, there is {a} narrow discussion about application providers' incentives from the perspective of governance. The following gives {a} detailed distribution of incentives in these blockchains.

In Bitcoin, software release notes of its official client, Bitcoin Core, record all contributors~\cite{BitcoinCore}. Transaction fees and mining rewards (i.e. new bitcoin tokens) are given to the corresponding validator (aka. miner)~\cite{Satoshi:bitcoin}. Node operators may be banned for a particular time period (by {default} 24 hours) if they {send} false information, as a waste of bandwidth and computing resources~\cite{bitcoin_developer_guide}. In Ethereum, {a} certain proportion of {the} total amount sold, which is about 60 million ethers, is distributed to compensate early contributors. Ethereum {applies} a similar incentive mechanism as Bitcoin: transaction fees and new ether tokens are both rewarded to validators~\cite{Ethereum_whitepaper}. In Dash, mining rewards are divided into three pieces: validators and masternodes get 45\% respectively, while the remaining 10\% is to fund improvement proposals~\cite{Dash_Core_developer}. Moreover, Dash implements a Quorum system {in which} a node will be deactivated after six violations~\cite{dash_whitepaper}. In Tezos blockchain, both block validators and endorsers (who confirm a block via signature) are incentivised by mining rewards along with a bond. Validators can receive the bond after a security cycle (i.e. one year). The bond will be forfeited if violations in their blocks are found (e.g. double signing). Inactive Tezos addresses are not allowed to validate blocks or vote for improvement proposals until they are {reactivated}~\cite{tezos_whitepaper}. In addition, Bitcoin and Dash both mention taxes for regulators~\cite{bitcoin_faq, dash_legal}, while Ethereum pays for audit services~\cite{ethereum_audit}. 

On the contrary, Hyperledger Fabric does not have a native cryptocurrency distribution, consequently, incentives in this blockchain platform refer to conventional business revenues. The project team are paid by application providers when IBM Cloud is utilised to deploy and maintain blockchain instances~\cite{fabric_pricing}, and the team will pay for audit services~\cite{fabric_audit}. Application providers, node operators and users of a Hyperledger Fabric instance belong to different organisations. These organisations establish business connections where blockchain is adopted. Hence, incentives of these stakeholders are based on institutional governance within a single organisation, or business agreements between multiple organisations.

\textbf{Analysis of Principle 3.}
The analysis of decision rights is dominantly about {the} project team, node operators, users and regulators. First, {the} project team is in charge of the design, development, and management {of} both {the} blockchain platform and off-chain community. In particular, {the} Hyperledger fabric project team provides consultation of blockchain services to their clients (i.e., application providers)~\cite{fabric_consult}. Secondly, node operators hold blockchain {replicas} locally, and block validators are responsible for the validation and confirmation of blocks. Dash introduces the concept of ``masternodes", who provide services to ensure the availability of the Dash blockchain~\cite{dash_whitepaper}. They are able to vote on governance and funding proposals. Thirdly, users submit transactions to use on-chains services. {They can also participate in governance decision-makings. For instance,} Ethereum users have cast votes {for} forking~\cite{carbonvote}. Finally, Bitcoin and Dash acknowledge that governments are making efforts to integrate {blockchain platforms} into formal {and} regulated financial systems~\cite{bitcoin_faq, dash_legal}. Ethereum and Hyperledger Fabric select third-party {authorities} to perform {audits}~\cite{ethereum_audit, fabric_audit}. Tezos mentions regulators for DApps~\cite{tezos_corporate_baking}. All {the} above descriptions of decision rights and decision-making processes can be found on official websites and documents to gain the trust of existing and possible stakeholders. However, limited information is provided about how application providers {are} involved in governance decisions, except {for} Hyperledger Fabric. In Hyperledger Fabric, application providers are significant for deciding whether to adopt blockchain technology, and also maintaining {an} off-chain community. Especially, they need {to} consider the actual business problems, overall ecosystem, business and governance model, and also legal issues before deployment~\cite{fabric_handbook}.

% \begin{itemize}
%     \item \textit{Principle 4: Establish role-based accountability through both institutional and technical means}
% \end{itemize}

%\textbf{Establishment explicit accountability.}
\textbf{Analysis of Principle 4.}
All selected blockchain platforms utilise both institutional and technical means for {the} accountability process. Institutional means include commit history in Github to locate the exact contributor on {a} specific update, and internal governance within {the} project team, as they all provide the respective institution structure on official websites. Technical means mainly rely on blockchain addresses of participants, and operation logs in {the} transaction information. Particularly, Dash develops a Proof of Service scoring system to recognise the contributions of masternodes~\cite{dash_readme}. Nevertheless, in the four permissionless public blockchains, it is still hard to identify on-chain participants for real-world accountability. Further investigation is required to analyse trade-offs between accountability and privacy. While in Hyperledger Fabric, transacting entities are not anonymous, all involved organisations and their employees are identified by a Certificate Authority associated with each organisation~\cite{fabric_doc}.
%Consequently, they are all marked ``partially implemented" on this principle.

% \begin{itemize}
%     \item \textit{Principle 5: Support ecosystem-level governance}
% \end{itemize}

%\textbf{Support ecosystem-level governance.} 
\textbf{Analysis of Principle 5.}
The results of Principle 5 are analysed regarding the four different layers as follows. %Note that for this principle, sufficiency of implemented governance also considered the coverage of lifecycle phases.

\textit{Data layer:} When submitting transactions, all blockchain platforms filter transactions that any included field violating the standard format or required software version will result in errors, and neither be accepted, broadcast, nor validated. Bitcoin allows multi-signature feature that a transaction is sent with {the approval} of a set of users~\cite{Bitcoin_for_business}. Further, it also enables ``CoinJoin" {which} aggregates the operations of several users into a single transaction~\cite{bitcoin_developer_guide}. Ethereum specifies oracle services of sending data to blockchain~\cite{Ethereum_Development_Documentation}.

After generation, transactions are kept in a pending pool for validation. Bitcoin states that Bitcoin Core clients keep pending transactions in non-persistent memory~\cite{bitcoin_developer_guide}. If a node is shut down, the memory pool will be lost. While in Dash, masternodes are randomly selected to relay the inputs, outputs and signatures of transaction sets~\cite{dash_whitepaper}. A transaction set collects multiple transactions of the same user, but the selected masternode does not know the exact on-chain identity. In a Hyperledger Fabric instance, a transaction needs to be endorsed by appointed nodes before the ordering service node packages transactions into blocks~\cite{fabric_doc}.

After validation, full node operators take the responsibility of storing the whole transaction history, while light nodes provide auxiliary transaction verification services by recording only block headers. Bitcoin implements Merkle tree and Bloom filter to test the membership of elements~\cite{bitcoin_developer_guide}, which can achieve data compression in light nodes. Ethereum also splits desired data into blocks with encryption techniques, and then builds up a Merkle tree~\cite{Ethereum_whitepaper}. Arweave's Proof of Access is applied to test whether a node maintains both {the} most recent and randomly past block~\cite{Ethereum_Development_Documentation}. Tezos utilises indexers to achieve {a} quick fetch of blockchain data~\cite{tezos_Indexers}. In addition, the current upgrade of Ethereum will create new chains as shards for data storage and improve transaction throughput~\cite{ethereum2}. Tezos is investigating zero-knowledge proof to ensure data privacy~\cite{tezos_privacy}. Hyperledger Fabric allows users to send private data, where original contents are transferred during the endorsement process, {and the} respective hash values are stored in blocks~\cite{fabric_doc}. When data is stored in blocks, the accessibility is set via on-chain smart contracts. Nevertheless, the removal of on-chain data is not explicitly covered, except that Ethereum mentioned a hard fork was conducted to solve the DAO attack, which reversed transactions within a specific time period~\cite{ethereum_history}.

%hence we marked ``partially implemented" for all four platforms.

%Validation of transactions relies on the deployed consensus mechanism to select a specific validator, who usually includes transactions with higher fees in the consequent block.

%Governance of on-chain data in these platforms lay emphasis over the transaction validation and confirmation, data storage and access control. Particularly, privacy is a critical issue that Bitcoin and Dash provide mechanisms to hide the sender and receiver of a transaction, Ethereum highlights that on-chain data are encrypted to build Merkle tree, and Tezos is investigating zero-knowledge proof to ensure data privacy. Nevertheless, the removal of on-chain data log is not explicitly covered in the documents.

\textit{Platform layer:} The critical difference of governance in planning, analysis and design phases is embodied in the choice of consensus mechanisms. Currently, Bitcoin, Ethereum and Dash all {implement} Proof of Work, while Tezos deploys Proof of Stake (PoS). However, Ethereum is undergoing an upgrade from PoW to PoS. Tezos validators need to hold a minimum of 8000 XTZ tokens~\cite{tezos_baking} while Ethereum's future PoS requires 32 ETH tokens~\cite{ethereum2}. Although {the} validator selection in Dash relies on PoW, becoming a masternode has a PoS-style demand for holding at least 1000 Dash tokens~\cite{dash_readme}. Hyperledger Fabric supports pluggable consensus protocols which can be customised according to application providers' requirements~\cite{fabric_doc}. Raft is recommended by {the} Hyperledger Fabric project team where a ``leader and follower" scheme is implemented. Note that block validators (aka. ordering service nodes) in Fabric instances are assigned by authorities instead of competed as in permissionless blockchains. In terms of platform implementation, Github is commonly leveraged to manage source code in these platforms, which can provide commit history to assist {the} off-chain accountability process. %In addition to the mainnet, they all provide testnet to run new functionalities, avoiding influences to the real-world business activities. 

Upgrading blockchain platforms require a complete process of submitting improvement proposal, analysis and conflict resolution via voting, and implementation of code. Ethereum adopts ``Carbonvote", which means stakeholders' votes are calculated according to their owned ether tokens~\cite{carbonvote}. In Dash, specific on-chain governance messages contain the proposals and corresponding votes. The acceptance of a proposal needs the approval of at least 10\% of the masternode network~\cite{dash_feature}. Improvements in Tezos are grouped {into} five periods: proposal, exploration vote, cooldown, promotion vote, and adoption, each phase lasts for about 14 days~\cite{tezos_governance}. When releasing updated code, accepted proposals are implemented via {a} hard fork in Bitcoin and Ethereum as they are less flexible, while the other two permissionless platforms have smoother transitions between different versions. Dash integrates updated code into {the current} platform but does not immediately active new features, until more than 80\% {of} participants update their clients~\cite{dash_feature}. Tezos's on-chain protocol has two parameters for version control, which can conveniently enable accepted proposals~\cite{tezos_whitepaper}. Upgrades in Hyperledger Fabric need to contemplate both each deployed instance and the overall platform. If an application provider intends to change the settings of a Fabric instance, it needs to generate a new configuration file, which requires signatures from other organisations in the same instance~\cite{fabric_doc}. For upgrading the overall Hyperledger Fabric platform, regular issues such as bug fixes and documentation improvements are accomplished via the normal GitHub pull request workflow, while major upgrades require consensus from {the} broader community~\cite{fabric_RFC}. As Hyperledger Fabric does not implement {an} inherent token distribution or universal blockchain instance, such decisions are finalised via off-chain approaches. For instance, contributors meetings are held to plan and review release progress, and discuss future directions~\cite{fabric_doc}.

Regarding the termination of blockchain, Ethereum proposes the concept of ``social contract", {in which entities} with {a} certain quantity of ethers can develop a candidate version of Ethereum~\cite{Ethereum_whitepaper}. The Hyperledger Foundation specifies {a project's} lifecycle, which includes ``deprecated" and ``end of life"~\cite{fabric_lifecycle}. Project maintainers should cast vote(s) on the deprecation proposal, and the Hyperledger Technical Steering Committee determines whether to stop {supporting} a project.

%Consequently, we marked Ethereum as ``implemented" and the other three are ``partially implemented".
%Although Ethereum covers a full lifecycle, we still mark it and the other three as ``partially implemented", as limited information about cross-chain interoperability can be found.

%while they all use Github to manage open-source code, and provide testnet to run new functionalities. Upgrading a blockchain platform requires voting on improvement proposals, and the implementation will cause hard fork in Bitcoin and Ethereum, while Dash and Tezos have smoother transitions between different versions. Regarding the termination of a blockchain platform, only Ethereum proposes the concept of ``social contract" that individual with certain quantity of ether can develop a candidate version of Ethereum. %Moreover, these blockchain platforms do not mention cross-chain interoperability in the documents

\textit{Application layer:} The intrinsic DeFi background fertilises cryptocurrency-related applications with different client software and exchanges in permissionless public blockchains. In addition, diverse DApps are built up based on all selected platforms. They all provide tutorials and APIs for DApp development to practitioners. Testnet can be {deployed} to run new functionalities, avoiding negative influences {on} real-world business processes. In particular, Ethereum notes that it does not attempt to limit its use in particular fields, while regulatory mechanisms should be designed to prevent harm~\cite{Ethereum_whitepaper}. Hyperledger Fabric enables {a} smart-contract level endorsement policy, which specifies nodes from certain organisations need to validate transactions related to particular smart contracts~\cite{fabric_doc}. 
%We marked this layer as ``partially implemented" for all four platforms as more specifications can be provided to cover the whole development lifecycle of DApps.

%In addition to the intrinsic DeFi background, diverse DApps are built up based on these platforms. Although the official documents have limited discussion on governance issues, they all provide tutorials for the development of DApps.

\textit{Community layer:} There {are} no specific onboarding or exit rules for permissionless blockchains, anyone can join or leave at any moment. The separation of community is based on stakeholders' different expertise and spotlights {on} the blockchain platform or applications. Every platform provides formal communication channels, such as online blogs, discussion {forums} (e.g., Stack Overflow) and social media, and also offline workshops and meetings. Regarding Hyperledger Fabric instances, each organisation has its own structure for governance and management of employees. In {the} Hyperledger project team, there are clear instructions on how to participate and contribute to the project~\cite{fabric_doc} and code of conduct~\cite{fabric_conduct}. In addition, we also note that Bitcoin, Ethereum, and Tezos all have their own \textit{benevolent dictator(s)} who have dominant decision rights on governance issues. In Bitcoin, Satoshi Nakamoto was regarded as the benevolent dictator of the Bitcoin ecosystem before his/her retirement~\cite{nakamoto}, while the co-founder of Ethereum, Vitalik Buterin, is still active in governance-related issues {in} the Ethereum ecosystem\footnote{Vitalik Buterin's website: \url{https://vitalik.ca/index.html}}. In Tezos, the foundation has a veto power for the first 12 months as a security measure~\cite{tezos_whitepaper}.
%With limited discussion and plan on the termination phase, we marked them all ``partially implemented" at this layer.

% \begin{itemize}
%     \item \textit{Principle 6: Manage legal and ethical conformation}
% \end{itemize}

%\textbf{Define legal compliance \& ethical responsibility.} 
\textbf{Analysis of Principle 6.}
Legal compliance and ethical responsibilities are heavily reliant upon the culture where a blockchain platform is deployed. Bitcoin mentions tax liability~\cite{bitcoin_faq} for individual incomes, and also lists that Bitcoin is now prohibited or restricted in certain areas~\cite{Cryptocurrency_Legality}. Ethereum discusses common scams to prevent serious risks~\cite{ethereum_scam}. Dash provides detailed clues about the conformation of Banking Secrecy Act, Anti-Money Laundering, and Know Your Customer regulations, and lists known scams, fake wallets and Ponzi or pyramid schemes {on} the platform~\cite{dash_legal}. Tezos specifies that a DApp ``CoinHouse" is under the highest legal standards in French law~\cite{tezos_corporate_baking}. Hyperledger Fabric suggests that application providers should contemplate legal issues, including data-sharing regulations, industry-specific regulations, intellectual property, commercial agreements, and local and international law, before adopting blockchain in their business models~\cite{fabric_handbook}. Specifically, Pal~\cite{fabric_gdpr_pal} and Smith~\cite{fabric_gdpr_smith} both discussed GDPR and Hyperledger Fabric. In addition, the Linux Foundation behind Hyperledger Fabric specifies that their Antitrust Policy complies with all applicable state and federal antitrust and trade regulation laws~\cite{fabric_antitrust}.

Ethical responsibilities found in these blockchain platforms mainly refer to transparency, data privacy, and integrity. Transparency and data integrity are achieved by the decentralisation level of permissionless public blockchains. On the other hand, data privacy is preserved by access control policies over {the} data layer. Note that Ethereum now is undergoing a vital upgrade which includes the replacement of Proof of Work {with} Proof of Stake, this change will enormously reduce the energy consumption and improve {the} sustainability of Ethereum~\cite{ethereum2}. In addition, Hyperledger Fabric highlights the need {for} trust as all involving stakeholders are identified.

%Finally, we marked ``partially implemented" for all four platforms in this principle, considering blockchain-related legal regulations and ethical principles are still under study and investigation.

% \begin{itemize}
%     \item \textit{Discussion}
% \end{itemize}

%To sum up, these four blockchain platforms have similar implementation of governance methods, while they differentiate with each other 

%\noindent \textbf{Discussion}

\textbf{Summary.} 
The case studies show that our proposed framework is applicable in real-world blockchain governance. All six high-level principles can be found in the five blockchain platforms. In addition, the case studies also {reveal} the similarities and differences across selected platforms, as well as some findings to improve their governance as follows. 

First, the major difference between these blockchains is how to distribute incentives and deployed consensus protocols. PoW and PoS are commonly applied, as they can choose validators who are regarded {as having} more contributions (either computation power or holding tokens), and hence should be granted more decision rights. {The choice of} consensus mechanisms is also dependent on the decentralisation level. For instance, Proof of Authority is suitable for {the} low decentralisation level while Proof of Elapsed Time can achieve {a} random selection of validators and build {a} highly decentralised environment. In addition, as permissionless public blockchains provide universal blockchain instances, their project teams are significant when arranging consensus and incentives. Consequently, the deviation of incentives reflects how {the} project team value the contribution of stakeholders in permissionless blockchains. In the selected four cases, {their} incentive mechanisms all involve a game theory to attract block validators, and preserve the fund for future development. From the perspective of DeFi, it is also leveraged to mint new tokens into the market. Incentives in permissioned public blockchains may rely on conventional business {models}, as they are usually adopted in various application scenarios where there is no inherent token distribution. In the case of Hyperledger Fabric, {the} project team is more like a service provider. All stakeholders are closely connected by business agreements.

Secondly, regarding the three governance dimensions (i.e., incentive, decision rights, and accountability), there is a lack of discussion on how application providers make governance-related decisions, and what incentives drive their behaviours in the four permissionless public blockchains. A possible reason is that these platforms intend to promote the development of decentralized autonomous applications and organisations where human interference is minimised to achieve on-chain autonomy. On the contrary, the significance of application providers is highlighted in permissioned blockchains, as they are the ones to decide whether to deploy a blockchain instance. In addition, there are different settings for node operators. For example, Dash specifies ``masternodes" as the on-chain committee dealing with governance issues like voting on improvement proposals. This may facilitate {the} future design of more special roles with certain rights to finalise governance decisions (e.g., transaction visibility). While in a permissioned public blockchain instance, on-chain roles are usually matched to corresponding off-chain positions. Further, different from mandatory identity verification in permissioned public blockchains like Hyperledger Fabric, permissionless public blockchains have an inherent weakness for on-chain accountability that individuals are hard to identify, which implies the low cost of violations. This requires investigation {of} digital identity, for instance, self-sovereign identity is a blockchain-based application, which may in turn help on-chain identification.

Thirdly, all lifecycle phases of each ecosystem layer are covered except for {the} termination {phase} in several selected cases. This may indicate the project teams' confidence that they can survive in the competitive market to some extent. We suggest a full development process, which considers the termination phase, can help enable a complete governance-driven architecture design. For example, data handling and transaction management require proper on-chain data removal mechanisms that respect data subjects' rights and comply with related regulations like GDPR.

The third remark further leads to our final finding that although blockchain-related legal compliance and ethical responsibilities are still under investigation, these platforms do provide related discussions and analysis. Future {studies} can be conducted to raise general awareness, and further integrate it with current blockchain governance methods. For instance, filters can be applied to examine data contents instead of merely transaction format, to prevent malicious information {from} being fed to blockchain.

Through the case studies, it is observed that the proposed framework can satisfy feasibility and applicability. For feasibility, we confirmed that the six principles in our framework are addressed and considered in {a} real-world context. In terms of applicability, we demonstrated how the proposed framework is applied to different blockchain platforms to scrutinise their governance structures. The results helped identify current gaps and limitations based on selected platforms.
% In addition, we did not extract other more high-level principles for blockchain governance.}

Hereby, we summarise several research directions to guide practitioners {in} the implementation of blockchain governance. Future study is required to translate the high-level principles into actionable mechanisms and patterns, explore the effects of architectural decisions on governance, and analyse corresponding trade-offs from the perspective of software engineering. For instance, design patterns can be extracted from extant blockchain platforms by scrutinising reusable governance solutions.

%{In addition, the evaluation also reflects certain limitations of our proposed framework. On one hand, the framework is not applied to analyse governance structures in permissioned private blockchains, which may influence generalisability of our framework. On the other hand, in order to help practitioners design and realise blockchain governance, further study is required to translate the high-level principles into actionable mechanisms and patterns from the perspective of software engineering.}

%the four platforms only provide tutorials to application providers for the adoption of blockchain in DApps, but there is a lack of discussion how they can participate in governance-related issues. 

%Through the case studies, it can be observed that the proposed framework can achieve feasibility and applicability. The framework itself can be applied to different blockchain platforms to analyse generalised governance methods. In addition, the evaluation reveals that selected cases all implement governance throughout their ecosystem respectively, and majority of elements and principles in our framework are addressed and considered in real-world context. The main differences between them are how to distribute incentives and the deployed consensus mechanisms. Nevertheless, limited discussion over application providers and arrangement for termination were found, and legal \& ethical conformance can be further strengthened in the documents.

%\begin{sidewaystable}[tbhp]
\begin{table*}[!tbhp]
\footnotesize
\centering
\caption{Comparison results of existing frameworks for blockchain governance.}
\label{tab:relatedwork}
%\begin{tabular}{p{0.12\textwidth}p{0.065\textwidth}p{0.068\textwidth}p{0.065\textwidth}p{0.065\textwidth}p{0.065\textwidth}p{0.065\textwidth}p{0.065\textwidth}p{0.065\textwidth}p{0.065\textwidth}}
\begin{tabular}{p{0.16\textwidth}p{0.052\textwidth}p{0.052\textwidth}p{0.052\textwidth}p{0.052\textwidth}p{0.052\textwidth}p{0.052\textwidth}p{0.052\textwidth}p{0.052\textwidth}p{0.052\textwidth}p{0.065\textwidth}}
\toprule
\multirow{3}{0.16\textwidth}{\bf{Governance aspects}} & 
\multirow{3}{0.052\textwidth}{Katina et al.~\cite{selected5}} &
\multirow{3}{0.052\textwidth}{Allen and Berg~\cite{selected7}} &
\multirow{3}{0.052\textwidth}{John and Pam~\cite{selected10}} &
\multirow{3}{0.052\textwidth}{Pelt et al.~\cite{selected11}} &
\multirow{3}{0.052\textwidth}{Beck et al.~\cite{selected14}} &
\multirow{3}{0.052\textwidth}{Howell et al.~\cite{selected15}} &
\multirow{3}{0.052\textwidth}{Werner et al.~\cite{selected31}} &
\multirow{3}{0.052\textwidth}{Hofman et al.~\cite{hofman2021blockchain}} &
\multirow{3}{0.052\textwidth}{Tan et al.~\cite{TAN2021101625}} &
\multirow{3}{0.065\textwidth}{Our framework}
\\ \\ \\
\midrule

Decentralisation level & $\times$ & $\circ$ & $\times$ & $\bullet$ & $\bullet$ & $\bullet$ & $\bullet$ & $\bullet$ & $\bullet$ & $\bullet$
\\
\cmidrule{1-11}

Stakeholders & $\circ$ & $\bullet$ & $\circ$ & $\bullet$ & $\bullet$ & $\bullet$  & $\bullet$ & $\bullet$ & $\bullet$ & $\bullet$
\\
\cmidrule{1-11}

Incentives & $\times$ & $\circ$ & $\circ$ & $\bullet$ & $\bullet$ & $\circ$ & $\bullet$ & $\bullet$ & $\circ$ & $\bullet$
\\
\cmidrule{1-11}

Decision rights & $\circ$ & $\circ$ & $\times$ & $\bullet$ & $\bullet$ & $\bullet$ & $\bullet$ & $\bullet$ & $\circ$ & $\bullet$
\\
\cmidrule{1-11}

Accountability & $\circ$ & $\times$ & $\circ$ & $\bullet$ & $\bullet$ & $\circ$ & $\circ$ & $\bullet$ & $\bullet$ & $\bullet$
\\
\cmidrule{1-11}

Ecosystem & $\circ$ & $\circ$ & $\circ$ & $\circ$ & $\circ$ & $\circ$ & $\circ$ & $\circ$ & $\circ$ & $\bullet$
\\
\cmidrule{1-11}

Lifecycle & $\circ$ & $\circ$ & $\circ$ & $\circ$ & $\circ$ & $\circ$ & $\times$ & $\circ$ & $\circ$ & $\bullet$
\\
\cmidrule{1-11}

Legal regulations & $\circ$ & $\bullet$ & $\circ$ & $\circ$ & $\circ$ & $\bullet$ & $\times$ & $\bullet$ & $\bullet$ & $\bullet$
\\
\cmidrule{1-11}

Ethical responsibilities & $\times$ & $\times$ & $\circ$ & $\times$ & $\times$ & $\times$ & $\times$ & $\circ$ & $\circ$ & $\bullet$ 
\\
\cmidrule{1-11}

\multicolumn{11}{r}{$\bullet$ Covered \quad \quad $\circ$ Partially Covered \quad \quad $\times$ Not Covered} \\

\bottomrule

\end{tabular}
\end{table*}
%\end{sidewaystable}

%\section{Related Work}
%\label{sec:relatedwork}
\subsection{Comparison}

We compared our proposed framework with existing governance frameworks for blockchain platforms. We used {the} identified governance aspects as comparison factors to evaluate different concerns of compared frameworks. ``Sufficiency" is used to examine whether the aspects are considered in the frameworks. A result is determined as ``not covered" ($\times$), ``partially covered" ($\circ$ ), or ``covered" ($\bullet$). The comparison results are illustrated in Table~\ref{tab:relatedwork}.

%propose seven interrelated elements (philosophy, theory, axiology, methodology, axiomatic, method and applications) for researchers studying in this topic, while Allen and Berg~\cite{selected7} provide a descriptive framework to understand exogenous and endogenous governance in blockchain. John and Pam~\cite{selected10} and Pelt et al.~\cite{selected11} both study on-chain and off-chain development processes to adopt governance. Beck et al.~\cite{selected14} formulate a blockchain governance framework which is centered with the three dimensions of decision rights, accountability, and incentives from IT governance. Howell et al.~\cite{selected15} focus on the membership and transacting relationships, and Werner et al.~\cite{selected31} develop a taxonomy of platform governance for blockchain. Hofman et al.~\cite{hofman2021blockchain} propose a high-level analytic framework, regarding six aspects (i.e. why, who, when, what, where, how) in the governance of blockchain. Tan et al.~\cite{TAN2021101625} analyse this research topic from the perspective of social sciences.

First, most compared frameworks involve consideration of decentralisation levels when either designing their frameworks or conducting evaluations as our framework. Nevertheless, \cite{selected7} limits its investigation to permissionless public blockchains, while \cite{selected5} and \cite{selected10} do not explicitly mention specific blockchain types. Existing frameworks highlight that the decentralisation {levels are} significant for governance structures, we further discuss how this aspect affects {the} following governance aspects in our work. For instance, we explain how incentives, decision rights, and accountability may be various in different blockchain types.

Secondly, existing frameworks are marked as ``covered" for stakeholders in blockchain governance, if at least two types {of} stakeholders are considered. Most existing frameworks place emphasis on {the} project team, node operator, and user, while analysis {of} application provider and regulator is limited. Notably, \cite{selected7} covers the same stakeholders as our framework. \cite{selected11, selected14, TAN2021101625} adopt decision rights, accountability and incentives from IT governance. In particular, \cite{selected14} and \cite{selected11} define multiple research questions regarding the three governance dimensions for further theoretical work. Compared to them, our framework illustrates a comprehensive mapping between five groups of stakeholders to the three governance dimensions of incentives, decision rights, and accountability. This mapping can help {the} broader community comprehend the authority, capability, and responsibility of different stakeholders for blockchain governance.

Further, most compared frameworks are marked as ``partially covered" for both ecosystem and lifecycle aspects. For {the blockchain} ecosystem, existing frameworks discuss the separation of on-chain and off-chain governance~\cite{selected7,selected11}, or focus on either side. In this paper, our framework provides a more extensive blockchain ecosystem with four refined layers (i.e., data, platform, application, and community), and discusses how governance can be implemented in each layer. Regarding when to implement blockchain governance, existing frameworks roughly divide the phases into exogenous and endogenous governance~\cite{selected7,selected15}. In the proposed framework, the governance process is elaborated in terms of each identified ecosystem layer. In particular, we adopt {the} software development lifecycle to describe the governance process of platform and application layers. Our framework specifies multiple constructive governance mechanisms regarding different lifecycle phases.

Finally, extant frameworks lack concrete instances of legal conformation. In our framework, \textit{Principle 6} is provided as a high-level guidance with discussion and multiple examples of both legal compliance and broader ethical responsibility.

It is observed in the comparison that there are significant gaps between compared frameworks and our proposed framework. This highlights the need for our framework again.

\section{Conclusion}
\label{sec:conclusion}

In this paper, we proposed six blockchain governance principles, and built a comprehensive framework to support better governance processes in blockchain and blockchain-based applications. The governance framework considers the level of decentralisation in different blockchain types, allocation of incentives, decision rights and accountability of stakeholders. Further, the proposed framework extends governance to the blockchain ecosystem, and highlights both legal compliance and ethical responsibilities in blockchain governance. We elucidated how governance is implemented for five blockchain platforms through case studies, identifying current gaps and limitations. Their open websites and documents confirm the feasibility and applicability of our proposed framework. Blockchain governance is an arduous and ongoing topic, and there is a continuing need of providing considered guidelines for the design and implementation of blockchain and applications. In the future, we plan to propose a design pattern catalogue, and explore architectural decisions for blockchain governance.


% \bibliographystyle{IEEEtran}
% \bibliography{bibliography}


\end{document}